# Classifying Songs with EEG

Prashant Lawhatre (prashant.lawhatre@msc2017.iitgn.ac.in)[1]
Bharatesh R Shiraguppi (bharatesh.rayappa@iitgn.ac.in)[1]
Esha Sharma (esha.sharma@iitgn.ac.in)[1]
Krishna Prasad Miyapuram (kprasad@iitgn.ac.in)[1]
Derek Lomas (J.D.Lomas@tudelft.nl)[2]

[1] Indian Institute of Technology Gandhinagar, India
[2] Delft University of Technology, Netherlands

**Abstract**

This research study aims to use machine learning methods to characterize the EEG response to music. Specifically, we investigate how resonance in the EEG response correlates with individual aesthetic enjoyment. Inspired by the notion of musical processing as "resonance", we hypothesize that the intensity of an aesthetic experience is based on the degree to which a participant's EEG entrains to the perceptual input.
To test this and other hypotheses, we have built an EEG dataset from 20 subjects listening to 12 two minute-long songs in random order. After preprocessing and feature construction, we used this dataset to train and test multiple machine learning models.

**Keywords:** EEG; Music; Synchrony; Machine Learning

## Introduction

To what extent does oscillatory activity in the human brain's EEG signal reflect oscillatory perceptual input, like music? Neural Resonance Theory (Large and Synder, 2009) models human beat perception in terms of the entrainment of neural oscillations to the rhythms of music. Over the past few years, researchers have documented extensive evidence of this neural entrainment of musical rhythms (Tal et al., 2017; Doelling and Poeppel, 2015). Further, researchers have documented the Frequency Following Response, which is the synchronization of neural frequencies to tonal frequencies, (Bidelman and Powers, 2018; Nozaradam, 2014). Some researchers have even been able to use oscillatory brain activity to reconstruct perceptual input (I. Kavasidis et al., 2017). Yet, all of these studies have required many repeated exposures to tones or rhythms in order to detect the oscillations in brain activity. Our current study sets out to design and validate machine learning (ML) models capable of accurately predicting which song a person is listening to, based on their brain waves (EEG), using only a single exposure to a 2 minute sample of the song.

## Related Works

Several researchers have built machine learning models to predict musical features in the OpenMIIR dataset (64 channel, 512hz). This dataset contains data from 10 subjects who listened to 12 music pieces (duration from 7 to 16 seconds) five separate times (Ntalampiras Stavros et al. 2019). Songs included titles like Mary Had a Little Lamb, the Star Wars Theme song and Jingle Bells.

Ntalampiras et al (2019) developed their models by training on 9 subjects and testing on the 10th subject. Their best per-subject classification rate was 50.8%, their worst was 38.3% and the overall average rate was 42.7%. The best per-song classification rate was 52.5% for Take Me Out to the Ball Game (no lyrics), while their worst was 24% for Chim Chim Cheree (no lyrics). When using their same approach using only musical signals, they achieved >99% classification accuracy.

Foster et al (2018) aimed to correlate Music Information Retrieval Features between music and EEG. They first extracted the features from the musical clip, then extracted the same features from the EEG data and finally conducted correlation tests between the two. The features, developed using the librosa library in Python, included the MFCCs (Mel Frequency Cepstral Coefficients), RMSE (Root Mean Square Error), spectral centroid, chroma STFT(Short Time Fourier Transform), spectral roll-off, tempogram, harmonics, and beats. Correlation tests included RSA (representational similarity analysis) and linear models 2 vs 2 tests. Only the MFCC and tempogram features had significant correlation. Using a logistic regression model, Foster et al (2018) achieved a classification accuracy of 0.287 on the testing dataset (which consisted of 20% of the listening events). The baseline random accuracy for this classification task would be 0.08 (1/12).

## Machine Learning Study Design

Our data collection protocol aims to support the development of machine learning models capable of predicting the music that a person was listening to during the EEG recording. Each unit or epoch of the EEG data is labeled with the music that was concurrently playing. In accordance with machine learning best practices, this labeled dataset is then divided into training and testing dataset. These data are used to develop and validate linear and non-linear models of the effects of music on the psychophysiological responses in the brain.

## Research Questions
1. To what degree can machine learning models predict enjoyment and familiarity of the song, based on the EEG data?
2. To what range can the machine learning models predict the song a person is listening to, based on the EEG data?
3. To what extent does neural resonance (EEG entrainment with musical features) support song prediction or the prediction of enjoyment/familiarity?
4. Which EEG metrics are most predictive?

## Hypotheses
1. Musical features found in the EEG data will provide predictive utility in the ML models.
2. Individuals that have a stronger positive aesthetic response to music (i.e., rate a song as more enjoyable) will exhibit a stronger resonance with the music (more brain areas synchronizing with the musical features), and therefore produce data that is more easily classified by the ML models. That is, we expect that more the more enjoyable songs will produce more predictive musical activity in the EEG (via Neural Resonance), resulting in greater predictive accuracy by the ML models.

## Data Collection
A participant is seated in a dimly lit room and provided instructions about the study. Following are the preparation of the EEG headset.
1. First we measure the circumference of the participants head to get the maximum possible size
2. Then select appropriate sized EGI net cap. KCl electrolyte solution is prepared in 1 ltr pure distilled water. Then the EGI net is immersed in that solution for 5 minutes
3. Reference electrode position is measured as the intersecting point on the lines between nasion (point in between eyebrows) and inion (middle point of skull ending at the backside) with preauricular points on both sides and then it is marked.
4. Then, the electrolyte soaked net is taken out and placed on the participant's head with placing the reference electrode first according the previously marked position
5. Finally the EEG net is connected to the amplifier box.
6. Each electrode is placed properly to bring the impedance within the acceptable range

Once everything is set, experiment begins with collecting basic information of age, gender and handedness. Then the participants were asked to read the instructions on the screen, which asks them to close their eyes after a single beep. That is followed by two minutes of silence, after that they listen to the first two-minute-long song, which is followed by a double beep as a signal to open eyes. They then rate "how much did they enjoy" the track they heard and how familiar the track is to them on a scale of 1-5. After the rating screen displays asking the participant to keep the eyes closed again. This continues for 10 seconds of silence before presenting the next track. Same procedure repeats until the last song has been rated. We then collect a final 2 minute period of silence with eyes closed. We collected data at 250 Hz and 1000Hz sampling rate on 128 channel EGI netstation system.

## Participants
20 participants were recruited from students, faculty and visitors of IIT-Gandhinagar. 16 participants were male, 4 female. All right handed with an average age of 25.3 years and standard deviation of 3.38.

## Data Analysis
Our formal data analysis methods can be described as having 4 phases: Preprocessing, Feature Development, Model Development and Testing models. In this paper, we are only reporting on an initial analysis procedure that used typical machine learning approaches with typical EEG features.

### 1. Pre-processing:
1. Capturing the music portions from the EEG data
2. Baseline correction (10,000 ms - 0 ms)
3. Filtering (cleanline) at 50Hz
4. Re referencing: average referencing with respect to 129 channel numbers
5. Bad channel rejection: Using automated option from EEGLAB toolbox, based on Probability, Kurtosis, Spectrum
6. Artifact removal using EEGLAB extension called "ADJUST"

### 2. Feature development:
Using preprocessed EEG data, this phase involved using multiple techniques to generate potentially predictive features. The feature development involves reducing the dimensionality of the 128 channel 1000 hz EEG into a far smaller set of values representing characteristics of the data epoch. For instance, in a single 10 second epoch, the alpha band power can be represented as a single number.

**The features used in our initial analysis included:**
1. Spectopo: band power of eeg alpha,beta,gamma,delta,theta
2. Wavedec: band power of eeg, using db8 kernel
3. DFA (dim, alpha): Detrended Fluctuation analysis(dim, alpha)
4. DFA (f): Detrended Fluctuation analysis(coeff.)
5. Entropy: Log Energy and Shannon Energy

### 3. Machine learning model training development:
1. After features extraction, models were trained
2. The features were also projected to lower dimension (for visualization purposes only) and were later used for training the above mentioned models

The below models were trained on segmented (10 second epochs) as well as unsegmented EEG signal (full 2 minute long epochs).

Table 1: Machine Learning models.

| Supervised learning | Unsupervised learning | Deep learning |
|---|---|---|
| AdaBoost | K-means clustering | CNN 1D |
| Decision tree | Hierarchical | CNN 2D |
| Gaussian process | Spectral | |
| Gradient boost | GMM | |
| KNN | | |
| MLP Naive bayes | | |
| Qda | | |
| Random forest | | |
| SVM | | |

### 4. Model testing:
In accordance with best practices in machine learning research, we set aside ⅓ of our data at the onset of this project. We randomly selected ⅓ of all epochs across all participant-song combinations, either from the 10 second long epochs or from two minute long epochs. We report on the performance of our developed models on the test dataset. This test data was used only once.

### Results
The top 3 performing algorithms. All used smaller 10-second long epochs, spectropo features and supervised models.

Table 2: Song classification accuracies of 3 classifiers.

| Model type | Model name | Features | Max song classification accuracy |
|---|---|---|---|
| Supevised_seg | KNN | Spectopo | 27.59 |
| Supervised_seg | Gradient Boost | Spectopo | 24.42 |
| Supervised_seg | MLP | Spectopo | 22.84 |

Figure 1: Confusion matrix based on top performing model, KNN with Spectropo

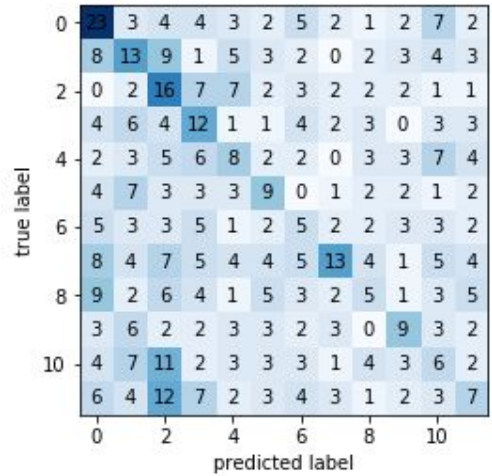

Confusion matrix for songs classification using spectopo feature. In the above matrix, 0 refers to song 1 and 11 refers to song 12

### Conclusion
Our study largely consists of building an EEG dataset that is densely labeled with the music playing at the time of the EEG recording. With 1000 samples per second and over 128 EEG channels, this produces millions of labeled data points for each individual subject. This labeled dataset lends itself to analysis using supervised machine learning models.

This is one of the preliminary attempts in creating a public dataset consisting of high-quality EEG data of diverse people listening to diverse music in a naturalistic way.

Our results validate several key assumptions about Neural Resonance Theory and raise new challenging questions for future research. Practically, our results inform new implications for the design of neurotechnologies and BCI. If oscillatory brain activity can be used to predict or reconstruct perceptual input, this could have significant applications. For instance, it might be possible to reconstruct imagined speech or music, directly from EEG.

## Acknowledgments

This work is financially supported by "Playpower Labs INC" and we would like to sincerely thank them.